\documentclass[pra,aps,twocolumn,showpacs]{revtex4}
\usepackage[pdftex]{graphicx}
\usepackage[]{times,amsmath,amsfonts,amssymb,epsfig,color}

\begin{document}

\title{Electromagnetic Continuum Induced Nonlinearity}

\author{Andrey B. Matsko}
\affiliation{OEwaves Inc., 465 North Halstead Street, Suite 140, Pasadena, CA 91107}

\author{Sergey P. Vyatchanin}
\affiliation{Faculty of Physics, Moscow State University, Moscow 119992 Russia}

\begin{abstract}
A nonrelativistic Hamiltonian describing interaction between a mechanical degree of freedom and radiation pressure is commonly used  as an ultimate tool for studying system behavior in opto-mechanics. This Hamiltonian is derived from the equation of motion of a mechanical degree of freedom and the optical wave equation with time-varying boundary conditions. We show that this approach is deficient for studying higher order nonlinear effects in an open resonant opto-mechanical system. Opto-mechanical interaction induces a large mechanical nonlinearity resulting from a strong dependence of the power of the light confined in the optical cavity on the mechanical degrees of freedom of the cavity due to coupling with electromagnetic continuum. This dissipative nonlinearity cannot be inferred from the standard Hamiltonian formalism.
\end{abstract}

\pacs{42.65.Sf,42.50.Wk}

\maketitle

\section{Introduction}

Opto-mechanics attracted a lot of attention as a tool for transferring purely theoretical quantum mechanical notion to experimental labs \cite{aspelmeyer14rmp}. Interaction of mechanical objects with light resulted in an efficient cooling of mechanical degrees of freedom \cite{kippenberg07oe,kleckner06n,Peterson2016} so single mechanical quanta became accessible. The light is able to manipulate the mechanical quanta, squeeze or entangle mechanical degrees of freedom \cite{mancini02prl,huang09njp,borkje11prl,akram13njp,liao14pra,woolley14pra}. Emission of coherent phonon radiation became possible \cite{grudinin10prl}.

Mechanical systems influence light as well, creating quantum entangled states between photons and phonons \cite{vitali07prl}. Quantum state transfer becomes possible between light and a mechanical system \cite{zhang03pra}. Finally, mechanical systems can modify the quantum properties of light, for instance, create squeezed light \cite{vyatchanin93jetp,mancini94pra,purdy12prx}.

Nonlinear physics also benefitted from the opto-mechanics \cite{Lemonde2013, Brawley2016, Peterson2016}. High spectral purity opto-mechanical oscillators were created \cite{carmon05prl,hosseinzadeh06pra}. Efficient optical frequency harmonics arising from the stimulated Brillouin scattering were used to generate narrow linewidth light \cite{li14ol} as well as low noise radio frequency signals \cite{li13nc}. Generation of a phonon frequency comb as well as mode locking of a mechanical distributed system was demonstrated \cite{savchenkov11ol}.

The beauty of an opto-mechanical interaction is in its clear physical picture based on Maxwell equations. Optical wave impinging on a mechanical object transfers its momentum to the object. Both cavity frequency and photon number changes as the result of such an interaction. Intricate physical phenomena can occur in the system if the mechanical body is moving fast, if it absorbs or scatters light, if its size is comparable with the optical wavelength, etc. However, the system simplifies significantly when optical photons confined in a closed (lossless) cavity interact with the nonrelativistic movable totally reflective cavity boundaries. Hamilton approach is usually applied to describe this kind of opto-mechanical interaction. In this paper using an example of a 1-D Fabry-Perot cavity we show that the Hamilton approach is deficient if one considers an externally pumped cavity. The energy exchange between the cavity and the optical pumping strongly depends on the position of the mirror, $x$, so the photon
number in the
optical mode changes significantly if the mirror motion is slow enough. This energy exchange dominates over high-order nonlinear by $x$ phenomena observed in the case of closed (lossless) optical cavity and this behavior cannot be predicted using a conventional opto-mechanical Hamiltonian. We show that the attenuation assisted nonlinearity can be so large, that high order mechanical harmonics can be readily generated in a mechanical system pumped with continuous wave light.

\section{Hamiltonian approach to opto-mechanics}

Interaction of a single optical mode and a single mechanical degree of freedom can be presented in quasistatic approximation in form \cite{mancini94pra,law95pra}
\begin{equation} \label{hint1}
H_{int}=-\hbar g \hat a^\dag \hat a \hat x,
\end{equation}
where $\hat a$ and $\hat a^\dag$ are photon creation and annihilation operators, $\hat x$ is a mechanical coordinate measured from the mechanical equilibrium point in the case of no light present, and $g$ is an opto-mechanical coupling constant. In the case of a 1D Fabry-Perot cavity with a movable mirror (Fig.~\ref{fig1}) this coupling constant is simply $\omega_0/L$ \cite{mancini94pra}, where $\omega_0$ is the carrier frequency of the light and $L$ is the distance between the mirrors. The photon number does not change in this case. Motion of the mirror results in change of the optical frequency.

We are interested in nonlinear behaviour of the opto-mechanical system and would like to derive a Hamiltonian that takes into account terms nonlinear in the mechanical coordinate $\hat x$. The expression (\ref{hint1}) directly follows from Maxwell equations. It is possible to write for electric field amplitude of light confined in an empty 1D Fabry-Perot cavity with totally reflecting mirrors (strictly speaking for steady state) the following equation
\begin{equation} \label{losslessdelay}
\hat a(t)= \hat a[t-2(L+ \hat x)/c],
\end{equation}
where $c$ is speed of light in the vacuum.

Assuming that $\hat x/L \ll 1$, in quasi-static approximation $L/c \gg \hat{\dot x}/x$, we can directly verify that expression
\begin{equation} \label{freq1}
\hat a= \hat a(0)e^{\pm i \pi c l t/(L +\hat x)},
\end{equation}
(where $l$ is the mode number) is a solution of Eq.~(\ref{losslessdelay}). The quasi-static approximation is needed to prohibit photon exchange between the modes and to require photon conservation in the mode. Since the system is unitary, this is equivalent to saying that the mode frequency depends on coordinate as $\pi c l/(L+\hat x)$ and the total Hamiltonian of a selected mode of the system is
\begin{equation}
\label{H4}
H= \hbar \omega_0 \hat a^\dag \hat a \frac{L}{L+\hat x},
\end{equation}
where the mode frequency is defined as $\omega_0=\pi c l_0/L$ ($l_0$ is integer, corresponding wavelength $\lambda_0=2 \pi/\omega_0$). The interaction Hamiltonian, defined as $H_{int}\equiv H-\hbar \omega_0 \hat a^\dag \hat a$, becomes
\begin{equation} \label{hint1a}
H_{int}=-\hbar g \hat a^\dag \hat a \frac{\hat x}{1+\hat x/L}.
\end{equation}
In general, this Hamiltonian should be utilized instead of (\ref{hint1}) to take the nonlinear terms $\hat x^n$ into account.

The Hamiltonian can be derived in a more explicit way using Eq.~\eqref{losslessdelay}. Introducing slow amplitude $\hat A$, so that $\hat a= \hat A\exp(-i\omega_0 t)$, we rewrite Eq.~(\ref{losslessdelay}) as
\begin{equation}
\label{A}
\hat A(t)=\hat A[t-2(L+ \hat x)/c] e^{4i \pi \hat x/\lambda_0}.
\end{equation}
The slow amplitude does not change much during the cavity round trip, which allows to use Taylor series
\begin{equation}
\hat A[t-2(L+ \hat x)/c] \simeq \hat A(t)- \frac{2}{c}(L+ \hat x)\hat{\dot A}
\end{equation}
to simplify Eq.~(\ref{A})
\begin{equation}
\hat{\dot A}\simeq \left( 1- e^{-4i \pi \hat x/\lambda_0}  \right ) \frac{c}{2(L+ \hat x)}
\end{equation}
or, for the case of small mechanical amplitude, $\lambda_0 \gg 4 \pi |\langle |\hat x| \rangle |$ ($\langle \dots \rangle$ stands for the expectation value), to a simpler differential equation
\begin{equation} \label{slowamp1}
\hat{\dot A}\simeq i\omega_0 A \frac{\hat x}{\hat x+L}.
\end{equation}
This equation is generated by Hamiltonian in the interaction picture
\begin{equation}
\tilde H_{int} = -\hbar \omega_0 A^\dag A \frac{\hat x}{L+\hat x},
\end{equation}
which is equivalent to Eq.~(\ref{hint1a}) if $g=\omega_0/L$.

The Hamiltonian (\ref{hint1a}) results in the equation for the mechanical degree of freedom
\begin{equation} \label{unitary1}
\ddot {\hat x}+ \omega_M^2 \left [ 1 + \alpha_{om1} \left (1-\frac{3}{2} \frac{\hat x}{L} + 2  \frac{\hat x^2}{L^2}   \right ) \right ] \hat x =\frac{\hbar g}{m}\hat a^\dag \hat a + \frac{F_s(t)}{m},
\end{equation}
where we truncated the nonlinear terms of the order higher than $(\hat x/L)^3$ and introduced an classical mechanical force $F_s(t)$; $m$ and $\omega_M$ are the mass and frequency of the mechanical system, respectively. To derive this equation we first differentiate Eq.~(\ref{hint1a}) by $\hat x$ and then decompose the result by powers of small parameter $\hat x/L$.

The nonlinearity of the system is defined by a dimensionless parameter
\begin{equation}
\label{alphaom1}
\alpha_{om1}= \frac{\hbar \omega_0 \hat a^\dag \hat a}{m\omega_M^2 L^2}.
\end{equation}
where we utilized $g=\omega_0/L$. The magnitude of $\alpha_{om1}$ is defined by the expectation value of the normalized DC shift of the mirror $\alpha_{om1} \sim 2 \langle \hat x \rangle /L \ll 1$.

Nonlinear terms appearing in Eq.~(\ref{hint1a}), $(\hat x/L)^n$, where $n>1$ is an integer, can result in generation of higher order mechanical harmonics if the size of the cavity is small enough. However, increase of the size to a kilometer range practically nullifies the effect. Moreover, the intrinsic mechanical nonlinearity of a micromechanical structure can be much larger if compared with the opto-mechanical part. For instance, similarly normalized mechanical nonlinearity parameter found from the Euler-Bernoulli theory applied to a micro-electro-mechanical system (MEMS) cantilever can exceed unity by an order of magnitude \cite{villanueva13prb,doolin2014pra,kaviani2015optica}. The lossless cantilever motion obeys to equation (please see \cite{villanueva13prb} for derivation)
\begin{equation} \label{mechanical}
\ddot x+ \omega_M^2 \left [ 1+ \frac{\beta_{geom}}{m \omega_M^2} \frac{ x^2}{{\cal L}^2}+ \frac{\beta_{iner}}{m \omega_M^2} \frac{ \dot x^2 + x \ddot x}{{\cal L}^2} \right ] x=\frac{F_s(t)}{m},
\end{equation}
where ${\cal L}$ is the cantilever length scaling in the micrometer range, $\beta_{geom}$ and $\beta_{iner}$ are geometrical and inertial nonlinear coefficients, respectively. It was shown that the effective dimensionless nonlinearity parameter $\alpha=(\beta_{geom}/(m \omega_M^2)-2\beta_{iner}/(3m)$ can exceed $-20$ for a real physical system. This is a much larger value if compared with the expected opto-mechanical nonlinearity $\alpha_{om1}$ involving reasonably small optical power. Therefore, it is reasonable to neglect by the ponderomotive mechanical nonlinearity in a unitary opto-mechanical system and consider only mechanical one.

\section{Open opto-mechanical system}

We found, though, that there is a dissipation-associated mechanism that results in several orders of magnitude increase of the light-mitigated mechanical nonlinearity. The effect has common features with additional rigidity arising in an opto-mechanical system when a mechanical degree of freedom  modulates the damping rate of a driven optical cavity \cite{mo09prl,elste09prl}. In what follows we derive the nonlinear terms using wave equation.
\begin{figure}[ht]
  \centering
  \includegraphics[width=6.cm]{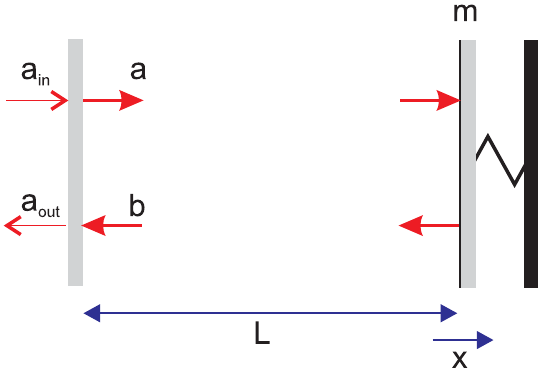}
\caption{  \small  Schematic of the 1D Fabry Perot resonator with movable mirror.} \label{fig1}
\end{figure}

Let us consider an empty 1D Fabry-Perot resonator pumped with a plain wave $a_{in}(X,t)=A_{in}(X,t) \exp [-i(\omega t-k X)]$, where $k=\omega/c$ is the wave vector and $X$ is the coordinate (Fig.~\ref{fig1}). The front mirror of the resonator, characterized with the power transmission $T$, is placed at position $X_1=0$, and the back, total, mirror is movable, so its coordinate becomes $X_2=L+\hat x(t)$, where $L$ is the distance between the mirrors and $\hat x(t)$ is the time dependent part of the total mirror coordinate. Standard equations describing electric field inside and outside of the resonator at the boundary of the input mirror ($X=0$) are
\begin{align}
\hat a(t)&= \sqrt{1-T} \hat b(t) +i\sqrt{T} \hat a_{in} (t), \label{inside}\\
\hat a_{out}& (t) = i\sqrt{T} b(t) + \sqrt{1-T} \hat  a_{in} (t)  , \label{outside} \\
\hat b(t) &\simeq \hat a\left (t-\frac{2[L+\hat x(t-L/c)]}{c} \right ) \left ( 1-2\frac{\hat {\dot x}(t)}{c} \right ) \label{feedback}.
\end{align}
Here term proportional to $\hat {\dot x}$ results from Doppler effect. While this term is usually small, it is necessary to keep it to sustain the right commutation relation for the coordinate and momentum of the mechanical system \cite{mzv96OC}.

Substituting Eq.~(\ref{feedback}) to Eq.~(\ref{inside}) we arrive to the equation for the field inside the resonator
\begin{align} \nonumber
\hat a- & \sqrt{1-T} \hat a\left (t-\frac{2[L+\hat x(t-L/c)]}{c} \right )
   \left ( 1-2\frac{\hat {\dot x}(t)}{c} \right ) \\
  &= i\sqrt{T} \hat a_{in}. \label{strict}
\end{align}
Equation~(\ref{strict}) coincides with Eq.~(\ref{losslessdelay}) for the nonrelativistic case and closed (lossless, $T \equiv 0$) resonator.

Equation~(\ref{strict}) has to be supplied with with equation for the coordinate of the movable mirror, that reads
\begin{align} \label{xdelayed}
\hat {\ddot x} (t)& + 2 \gamma_M \hat {\dot x}+\omega_M^2  \hat x(t) = \frac{\hbar \omega_0}{2 m L} \times \\
  \nonumber \times&  \left [ \hat a^\dag \left (t-\frac{L+\hat x(t)}{c} \right )
    \hat a \left (t-\frac{L+\hat x(t)}{c} \right ) + \right.\\
    \nonumber  & + \left. \hat b^\dag \left (t+\frac{L+\hat x(t)}{c} \right )\hat b \left (t+\frac{L+ \hat x(t)}{c} \right ) \right ] +\frac{F_s(t)}{m}.
\end{align}
Here force $F_s(t)$ includes both the signal and Langevin terms; mechanical attenuation $\gamma_M$ is small.

There are two cases when set (\ref{strict}) and (\ref{xdelayed}) can be simplified: $\omega_M L/c <<1$ and $\omega_M L/c =\pi j$, where $j$ is a natural number. In the first case the opto-mechanical interaction results in generation of optical harmonics localized within single optical mode. In the second case the mechanical frequency corresponds to the free spectral range of the resonator, so several optical modes (optical frequency comb) are generated due to the opto-mechanical interaction.

In the case of short enough optical cavity ($\hat a \gg L\dot{\hat a}/c$) we derive from Eq.~(\ref{strict}) a simplified equation for the slow intracavity field amplitude
\begin{align}\label{atrans}
\hat {\dot a}+ \left [\Gamma(\hat x)-i\Delta(\hat x) \right ] \hat a = \frac{i\sqrt{T}}{\tau} \hat A_{in},
\end{align}
where the coordinate-dependent optical attenuation and dispersion are given by formulas
\begin{align}
\Gamma(\hat x) & = \frac{1}{\tau} (1-\sqrt{1-T}\cos [2k(L+\hat x(t))]), \\
\Delta(\hat x) & =\frac{1}{\tau} \sqrt{1-T}\sin [2k(L+\hat x(t))],
\end{align}
$\tau=2L/c$ is the cavity round trip time. Neglecting by the small terms associated with the Doppler effect as well as assuming $\omega_M \tau <<1$ we also simplify the equation (\ref{xdelayed}) for the mechanical system
\begin{align} \label{xdelayed1}
\hat {\ddot x}  + 2 \gamma_M \hat {\dot x} + \omega_M^2  \hat x = \frac{\hbar \omega_0}{ m L}  \hat a^\dag  (t  )\hat a (t ) +\frac{F_s(t)}{m}.
\end{align}

To solve this set of equations we assume that $F_s(t)$ is small and look for the solution in the vicinity of steady state defined by expectation values for the field and mechanical amplitudes
\begin{align}
A & \simeq \frac{i\sqrt{T}}{\tau} \frac{\hat A_{in}}{\Gamma_0-i\Delta_0}, \\
x_0& \simeq \frac{\hbar \omega_0}{ m \omega_M^2 L} |A|^2,
\end{align}
where
\begin{align}
\Gamma_0 &= \frac{1}{\tau} (1-\sqrt{1-T}\cos [2k(L+x_0)]), \\
\Delta_0 &=\frac{1}{\tau} \sqrt{1-T}\sin [2k(L+x_0)].
\end{align}
It is also assumed for convenience that $x_0$ includes all the smaller order DC terms appearing during the analysis of the nonlinear system. In the following analysis we consider only time dependent part of coordinate $\delta \hat x = \hat x-x_0$.

General analysis of the opto-mechanical system is rather involved. We are interested in evaluation of the nonlinear response and consider the exact resonant case ($\Delta_0=0$). We formally solve Eq.~(\ref{atrans}) for the field amplitude and substitute the solution into Eq.~(\ref{xdelayed}). Linear in the coordinate terms responsible for the well known ponderomotive attenuation and rigidity disappear for the resonant tuning of the pump light. The cubic nonlinearity terms also proportional to the optical detuning disappear as well. Only quadratic in coordinate terms survive.

The nonlinear equation for the mechanical degree of freedom with excluded optical variables can be presented in form
\begin{align}  \label{deltax2a}
\delta \hat {\ddot x}  + 2 \gamma_M \delta \hat {\dot x} + \omega_M^2 \left [ 1+\alpha_{om2}  \frac{\delta \hat x}{L} \right ] \delta \hat x  & =\frac{F_s}{m},
\end{align}
where the dimensionless quadratic nonlinearity parameter $\alpha_{om2}$ depends on the frequency of the forced oscillation. For instance, for the case of resonant mechanical force ($F_s=f_s\cos \omega_M t$) and relatively low quality factor of the optical cavity ($\Gamma_0 \gg \omega_M$) the nonlinearity parameter is
\begin{equation}
\label{om2}
   \alpha_{om2}= -4\,\frac{\hbar \omega_0 |A|^2}{m\omega_M^2} \frac{Q^2}{L^2}.
\end{equation}
It is obtained using expression $Q=\omega_0/(2 \Gamma_0)$ for the optical quality factor.

The equation (\ref{unitary1}) for the mechanical coordinate obtained for the closed (unitary) opto-mechanical system, also contains a quadratic term $\alpha_{om1}$ \eqref{alphaom1} which is $4Q^2 \gg 1$ times smaller than  $\alpha_{om2}$. Therefore, to find the nonlinearity in a correct way the unitary model has to be adjusted.

An approximate solution of the equation with respect to the expectation value of coordinate is
\begin{align}  \label{deltax2c}
\delta x& \simeq \frac{f_s}{2 m \gamma_M \omega_M} \sin (\omega_M t) + \\
   & \frac{\alpha_{om2}}{L} \left (  \frac{f_s}{2 m \gamma_M \omega_M} \right )^2 \cos (2 \omega_M t),
   \nonumber
\end{align}
where we omitted the zero frequency term assuming it to be a part of $x_0$. Equation (\ref{deltax2c}) shows that analysis of the mechanical spectrum allows evaluating the opto-mechanical nonlinearity.

For some practical applications it is useful to consider the case of high frequency force $F_s=f_s \cos(\omega_f t)$,  where $\omega_f \gg  \omega_m$, but $\omega_f \tau \ll 1$. In this case the nonlinearity reduces, but still is large
\begin{align}
   \label{deltaxFM}
   \delta x & \simeq - \frac{f_s}{ m \omega_f^2} \sin (\omega_f t) + \\
      & +  \frac{\omega_M^2}{16 \omega_f^2L}
	 \left[ \alpha_{om2}^\text{fm}e^{2 i\omega_f t} + \alpha_{om2}^{\text{fm}*}e^{-2 i\omega_f t}\right]
	 \left (  \frac{f_s}{ m \omega_f^2} \right )^2 ,\nonumber\\
\label{om2FM}
   \alpha_{om2}^\text{fm}& = 4 \, \frac{\hbar \omega_0 |A|^2}{m\omega_M^2} \frac{Q^2}{L^2}\, S,\quad
      S= \frac{-\Gamma_0^3}{(\Gamma+i\omega_f)^2(\Gamma_0 + 2i\omega_f)}.
\end{align}

Presence of the strong quadratic opto-mechanical nonlinearity contrasts with the absence of the similar term in the purely mechanical nonlinearity of the system. The physical nature of this opto-mechanical nonlinearity is related to the reduction of the intracavity power when the system deviates from the optical resonance. The power drops independently on the direction of the mechanical motion.

The pure mechanical nonlinearity is of cubic nature (Eq.~\ref{mechanical}). The nonlinearity of the unitary system  contain a small cubic part  $2\alpha_{om1}$  for the normalization selected in Eq.~\ref{unitary1}. The cubic nonlinearity terms are also introduced to the open opto-mechanical system for $\Delta_0 \sim \Gamma_0$. Omitting lengthy derivations we write for the corresponding cubic nonlinear coefficient
\begin{equation}
\alpha_{om3} \simeq \frac{k^3L\hbar \omega_0 |A|^2}{m\omega_M^2 T^3}.
\end{equation}

It is easy to see that this nonlinearity is $k^3 L^3/T^3 \gg 1$ times larger than the nonlinearity $\alpha_{om1}$ of the optically closed (lossless) opto-mechanical system.   The reason for the nonlinearity enhancement is again the interaction of the opto-mechanical system with continuum resulting in the change of the optical power in the cavity when the position of the mirror changes.

The magnitude $\alpha_{om3}$ can exceed the unity and be comparable with MEMS nonlinearity parameter $\alpha$ for a small number of optical photons in the cavity. Really, for an opto-mechanical system with MEMS mirror we get $\alpha_{om3} \simeq 10^3$ for $\lambda=532$~nm, $L=0.1$~cm, $|A|^2=10^2$, $m=1$~mg, $\omega_M=2\pi \times 1$~MHz, and $T=10^{-3}$.

The results of our calculations have qualitative match with experimental data. Opto-mechanical systems used to demonstrate generation of multiple equidistant optical harmonics separated by the mechanical frequency. The neighboring harmonics are approximately of the same magnitude. It means that the system has both strong odd and even nonlinear terms. Pure mechanical nonlinearity tends to have mostly odd terms. Presence of even terms is also possible if the system is prestressed, however their magnitude is usually small. Presence of the significant quadratic nonlinearity of the opto-mechanical system explains observed experimentally efficient generation of the optical sidebands separated from the pump carrier by the doubled mechanical frequency.

\section{Free mass interferometer}

It is interesting to estimate the opto-mechanical nonlinearity in the case of  $\omega_M \rightarrow 0$ since the nonlinearity increases with $\omega_M$ decrease.  Such a configuration is practically realized in the Advanced Laser Interferometric Gravitational Observatory (aLIGO) \cite{AdvLIGO2015, AdvLIGO2010} which can be reduced to an equivalent 1D Fabry-Perot cavity \cite{Buonanno2003} (corresponding to so called signal recycling mode) with movable mirror having mass $m=10$~kg and frequency $\omega_M/2\pi \sim 0.1$~Hz. The bandwidth of the optical cavity is about working bandwidth of aLIGO (it  is varied by position of signal recycling mirror), in estimates below we assume $\Gamma_0/2 \pi \sim 300$~Hz.  The zero and first order opto-mechanical effects are very important in this case. The zero order pondoromotive effect associated with the radiation pressure results in accelerated motion of the mirror that cannot be tolerated. To handle this effect, an electronic feedback is involved \cite{Barsotti2010,
AdvLIGO2015}. Because of the feedback loops the opto-mechanical system cannot be
considered using the simplest model presented above, however the mirror can be treated as a free mass in 30-1,000~Hz frequency range.

We can use Eq.~(\ref{om2FM}) to evaluate nonlinearity in this case for LIGO parameters \cite{AdvLIGO2015}.  Selecting $\Gamma_0= 2 \pi \times 300$~rad$/$s, $\omega_f=2 \pi \times 10^2$~rad$/$s, $P=800$~kW, $m=10$~kg (reduced mass), $L=4$~km, $\hbar \omega_0 |A|^2=2LP/c$, $\lambda= 1064$~nm, we arrive at
\begin{align}
\label{EstLigo}
  \frac{|\alpha_{om2}^\text{fm}|\omega_M^2}{8 \omega_f^2  L} & = Q^2\,
    \left(\frac{P}{m \omega_f^2L^2 c}\right) |S| \simeq  8\cdot 10^6\text{~m}^{-1}\,. % \\
%     \frac{|\alpha_{om2}^\text{fm}|\omega_M^2}{8 \omega_f^2  L} &=
%     4 \, \frac{P\cdot 2L/c}{m\omega_M^2} \frac{Q^2}{L^2}\, |S|\cdot \frac{\omega_M^2}{8 \omega_f^2  L}
%     \Rightarrow \eqref{EstLigo}
    \end{align}

In other words, if the magnitude of the first mechanical harmonic is $f_s/(m \omega_f^2) = 0.01$~nm, the magnitude of the second mechanical harmonic is about $8\times 10^{-16}$~m.

This can be easily detected in Advanced LIGO \cite{AdvLIGO2015, AdvLIGO2010,Martynov2016}. The unitary model predicts the magnitude to be many orders of magnitude smaller, which is practically undetectable in the system.

\section{Conclusion}

In this paper we have shown theoretically that opto-mechanical nonlinearity induced due to the open nature of the system can be much larger if compared with the nonlinearity of a optically closed (lossless) opto-mechanical system having the same other parameters. The effect arises due to the variation of the intracavity photon number in the open system as a function of the mechanical coordinate. In contract, the photon number of the lossless opto-mechanical system is conserved and only the frequency of the photons change due to variations of the mechanical degree of freedom. We found that the mechanical nonlinearity induced by the optical degree of freedom can be comparable with purely mechanical nonlinearity both in small scale for micro-mechanical cantilevers and in large scale for 40 kg free masses (mirrors) in Advanced LIGO interferometer.

\acknowledgments

S.V. acknowledges support from  Russian Science Foundation (Grant No. 17-12-01095, researches on Sec.IV) and National Science Foundation (partially, Grant No. PHY-130586).

\end{document}